\documentclass{article}

\usepackage{arxiv}

\usepackage[utf8]{inputenc} 
\usepackage[T1]{fontenc}    
\usepackage{hyperref}       
\usepackage{url}            
\usepackage{booktabs}       
\usepackage{amsfonts}       
\usepackage{nicefrac}       
\usepackage{microtype}      
\usepackage{lipsum}		
\usepackage{graphicx}
\usepackage{cite}
\usepackage[version=4]{mhchem}
\graphicspath{ {./images/} }

\title{Dynamics of Aggregation Processes and Electrophysical Properties of Transformer Oil-Based Magnetic Fluids}

\author{ \href{https://orcid.org/0000-0001-8915-2411}{\includegraphics[scale=0.06]{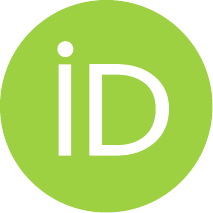}\hspace{1mm}Alexander D.~Kurilov}\\
	Federal State University of Education\\
	105005, Russia, Moscow, 10a Radio St. \\
    Prokhorov General Physics Institute\\
    of the Russian Academy of Sciences\\
    119991, Russia, Moscow, 38 Vavilova St.\\
	\texttt{ad.kurilov@gmail.com} \\
	\And
	Anastasia V.~Gubareva \\
	Federal State University of Education\\
	105005, Russia, Moscow, 10a Radio St. \\
    \And
	Sergei A.~Zubkov \\
	Federal State University of Education\\
	105005, Russia, Moscow, 10a Radio St. \\
    \And
    Yulia A.~Alekhina \\
    Lomonosov Moscow State University \\
    119991, Russia, Moscow, Leninskie Gory 1-2 \\
    \And
    Alexander V.~Simakin \\
    Prokhorov General Physics Institute\\
    of the Russian Academy of Sciences\\
    119991, Russia, Moscow, 38 Vavilova St.\\
    \And
	Denis N.~Chausov \\
	Prokhorov General Physics Institute\\
    of the Russian Academy of Sciences\\
    119991, Russia, Moscow, 38 Vavilova St.\\
}

\begin{document}
\maketitle

\begin{abstract}
Magnetic fluids exhibit tunable structures and electrophysical properties, making them promising for adaptive optical systems, biomedical sensors, and microelectromechanical devices. However, the dynamic evolution of their microstructure under varying magnetic fields remains insufficiently explored.

This study investigates the structural and dielectric properties of transformer oil-based magnetic fluids containing 0.2--10 vol\% magnetite nanoparticles, across a frequency range of 20 Hz to 10 MHz. Particular attention is given to the dynamics of aggregate reorientation in response to alternating magnetic fields. Experimental results demonstrate that low nanoparticle concentrations lead to a linear increase in dielectric permittivity and conductivity, consistent with the Maxwell-Wagner model. In contrast, higher concentrations exhibit conductivity saturation and dispersion effects due to the formation of elongated aggregates.

An analysis based on the Boyle polarization model describes the relaxation and structural changes associated with aggregation dynamics. Changes in the magnetic field orientation induce aggregate reconfiguration and significant structural transformations. At early stages, elongated chains form, subsequently thickening until an equilibrium state is reached. Elevated temperatures accelerate these processes by reducing medium viscosity and aggregate order.

The findings highlight the critical role of reorientation dynamics in designing high-speed magnetic sensors, vibration isolation systems, and adaptive devices operating in dynamic magnetic environments.
\end{abstract}


\section{Introduction}
Magnetic fluids are colloidal suspensions of magnetic nanoparticles stabilized by surfactants in a carrier liquid. These systems possess unique properties, including the ability to dynamically alter their structure and macroscopic characteristics in response to external magnetic fields. As a result, magnetic fluids find applications across a wide range of fields, from magnetorheological devices and sensors to electronics and biomedical technologies \cite{kole2021engineering, socoliuc2022ferrofluids, philip2023magnetic, oehlsen2022approaches}.

Among these applications, particular attention has been paid to the dielectric properties of magnetic fluids in variable electric and magnetic fields, as these investigations offer critical insights into the mechanisms of polarization, relaxation, and energy dissipation in such systems \cite{malaescu2002dielectric, batalioto2020electric, malaescu2023investigations}. Dielectric spectroscopy is a powerful tool for exploring the frequency-dependent behavior of magnetic fluids. This technique is highly sensitive to polarization and relaxation processes arising from interactions among particles, their aggregates, and the carrier medium \cite{batalioto2024impedance, rajnak2020controllability, rajvnak2023dielectric}.

Magnetic fluids exhibit a complex interplay of factors influencing their dielectric response, including nanoparticle concentration, aggregation processes, and the anisometric nature of their internal microstructure. For instance, the formation of chains or clusters of magnetic nanoparticles under a magnetic field induces dielectric anisotropy, which depends on the field’s strength and orientation. As the field intensity increases, the system transitions from isolated particles to microclusters, and ultimately to elongated structures. This induced dielectric anisotropy is driven by differences in polarization mechanisms along and perpendicular to the field direction, as well as charge transfer between particles within clusters \cite{malaescu2023investigations, demin2021polarization}.

The formation of anisotropic structures in magnetic fluids and their impact on dielectric properties have been well described. Previous studies have highlighted the role of dipole-dipole interactions in forming elongated aggregates under magnetic fields, significantly altering the effective dielectric parameters of magnetic fluids \cite{espurz1989magnetically, rajnak2014dielectric}. These effects are particularly pronounced in systems with high particle concentrations, where interactions between aggregates amplify the anisotropy of dielectric properties.

Induced dielectric anisotropy is strongly frequency-dependent and is associated with two main mechanisms: interfacial polarization (Maxwell-Wagner polarization \cite{li2021direct, gao2023charges}) and hopping ionic conductivity \cite{di2012broadband, pelster2004microstructure}. Both mechanisms dominate at low frequencies and are enhanced in systems with high conductivity contrast between the particles and the carrier medium. At higher frequencies (above 1 GHz), interfacial polarization occurs at the level of individual nanoparticles. Therefore, broadband dielectric spectroscopy is the most accurate method to capture structural changes within the system as a function of concentration and magnetic field direction, consistent with effective medium approximations such as the Bruggeman model \cite{pelster2004microstructure, spanoudaki2002frequency, lesmes2001dielectric, ji2019variational}.

Subsequent studies expanded these findings by incorporating models that relate cluster geometry to the frequency-dependent anisotropy of dielectric properties (e.g., Hanai’s model) \cite{adohi2008application, mansoorifar2017accuracy}. These results underscore the importance of considering cluster anisometry and orientation for accurately describing the dielectric properties of magnetic fluids. Cluster stability is determined not only by dipole-dipole interactions but also by van der Waals forces at small particle separations (approximately 1 nm) \cite{pelster2004microstructure}.

Theoretical approaches to describing these systems have evolved over time, incorporating factors such as particle shape distribution, material properties, and the presence of nanoparticle coatings. For instance, an approach developed for studying biological systems using dielectric spectroscopy has been applied to analyze membranes, liposomes, cells, and microcapsules \cite{asami2002characterization}.

A crucial factor in low-frequency relaxation for dispersed systems is the electric double layer (EDL), which forms around charged nanoparticles \cite{dong2017insight}. The EDL comprises a compact layer of counterions and a diffuse outer layer, both of which contribute to the overall dielectric response. Interfacial polarization mechanisms within clusters must be accounted for, especially at low frequencies, where EDL polarization becomes prominent \cite{rajnak2014dielectric, joseph2023complex, zhao2006dielectric, singh2014theory}.

While many studies have focused on static or equilibrium structures, the dynamics of reorientation and microstructural evolution under changing magnetic field directions have received less attention. This study addresses this gap by examining the dielectric properties of transformer oil-based magnetic fluids over a wide concentration range (0.2–10 vol\%) with a focus on reorientation dynamics and microstructural formation immediately following changes in magnetic field direction. Measurements were conducted across a frequency range of 20 Hz to 10 MHz, capturing both low-frequency interfacial polarization effects and EDL contributions. High-frequency parasitic effects were carefully mitigated to ensure measurement accuracy \cite{chassagne2016compensating}.

To interpret the observed phenomena, this study employs the Boyle polarization model \cite{boyle1985electrical}, which provides deeper insights into microstructural formation, relaxation processes, and dielectric anisotropy in high-concentration magnetic fluids.

\section{Materials and Methods}
\subsection{Magnetic Fluid Samples}
Transformer oil-based magnetic fluid (MFTO) samples with magnetite (\ce{Fe3O4}) nanoparticles were synthesized by the Problem Research Laboratory of Applied Ferrohydrodynamics at Ivanovo State Power Engineering University. The synthesis employed chemical precipitation from a liquid phase in the presence of a surfactant, with oleic acid used as a stabilizer. The magnetite concentration in the samples ranged from \(0\) to \(10 \, \text{vol\%}\).

\subsection{Magnetic Properties}
The magnetic properties of the fluids were studied using a Lake Shore 7410 vibrating sample magnetometer. Measurements were conducted at room temperature across a field range of \(-800 \, \text{kA/m}\) to \(800 \, \text{kA/m}\). The resulting magnetization curves were analyzed to determine magnetic susceptibility, saturation magnetization, and to perform magnetogranulometric analysis.

\subsection{Density Measurements}
Sample density was measured using the pycnometric method over a temperature range of \(263 \, \text{K}\) to \(323 \, \text{K}\). A \(5 \, \text{mL}\) glass pycnometer was immersed in a liquid cryothermostat (LOIP FT-316-25). Following temperature stabilization to within \( \pm 0.1 \, \text{K} \), the pycnometer was weighed on an analytical balance (Gosmetr VL-220M) with a precision of \(10^{-4} \, \text{g}\).

\subsection{Transmission Electron Microscopy (TEM)}
Morphological analysis was performed using a LIBRA 200 FE HR transmission electron microscope (Carl Zeiss AG) at an accelerating voltage of \(200 \, \text{kV}\) and a resolution of \(0.24 \, \text{nm}\). Sample preparation involved diluting the magnetic fluid with pure transformer oil, precipitating the particles, and repeatedly washing with dimethylformamide (DMF) before reprecipitation. A drop of the resulting solution was placed on a TEM grid and air-dried. The average particle size calculated from TEM images was \(16 \, \text{nm}\).

\subsection{Dielectric Spectroscopy}
The dielectric properties of the magnetic fluids were investigated using a precision impedance analyzer (WK65120P) across a frequency range of \(10^{1}\)–\(10^{8} \, \text{Hz}\). To avoid nonlinear and orientational effects, the test signal voltage was limited to \(0.5 \, \text{V}\). The measurement cell consisted of a parallel-plate capacitor filled with the sample under study.

During experiments, the capacitance and conductivity values were recorded to calculate the real and imaginary components of the dielectric permittivity \cite{barsoukov2018impedance}:
\begin{equation*}
	\varepsilon' = \dfrac{C - C_{p}}{C_{0} - C_{p}}, \quad \varepsilon'' = \dfrac{G}{\omega(C_{0} - C_{p})},
\end{equation*}
where \(C_{0}\) is the capacitance of the empty cell, \(C\) is the capacitance of the filled cell, \(C_{p}\) is the parasitic capacitance, \(G\) is the conductivity, and \(\omega = 2\pi f\) is the angular frequency of the test signal.

The filled cell was sealed and placed in a thermostatic circuit with temperature controlled by a flow thermostat (LOIP FT-316-25) over a range of \(273 \, \text{K}\) to \(323 \, \text{K}\). Each sample was thermostated for at least 10 minutes before measurements, ensuring temperature stabilization to \( \pm 0.1 \, \text{K} \).

To minimize high-frequency distortions, a compensation method for parasitic inductance and resistance effects was applied, as described in previous studies \cite{perkowski2012dielectric, perkowski2021parasitic}. Dielectric spectra of magnetized magnetic fluid samples were recorded after preconditioning in a magnetic field with a strength of \(400 \, \text{kA/m}\) for at least 48 hours, ensuring equilibrium aggregation states.

Measurements were automated using custom Python-based software, providing high accuracy and reproducibility. For complete spectral analysis, 100 points were evenly distributed on a logarithmic scale across the frequency range of \(20 \, \text{Hz}\) to \(120 \, \text{MHz}\). For dynamic measurements of time-dependent characteristics, fixed frequencies of \(20 \, \text{Hz}\), \(100 \, \text{Hz}\), \(1 \, \text{kHz}\), \(10 \, \text{kHz}\), \(100 \, \text{kHz}\), and \(1 \, \text{MHz}\) were used \cite{kurilov2024dynamic}.

\section{Results and discussion}
\subsection{General Characterization of Samples}
\begin{figure}[h!]
\centering
\includegraphics[width=0.75\linewidth]{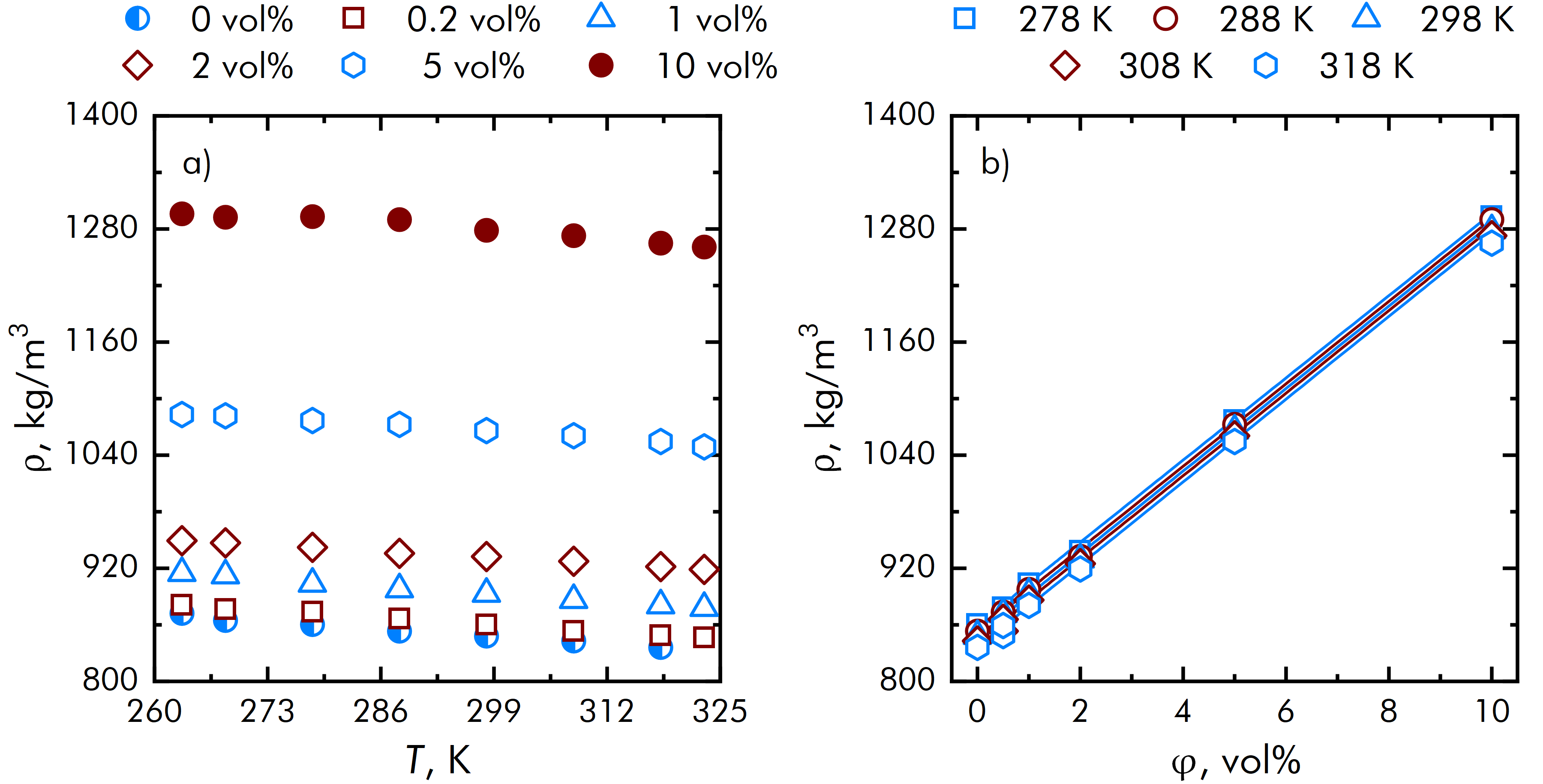}
\caption{Temperature-dependent (a) and concentration-dependent (b) density of magnetic fluid samples.}
\label{fig:fig1}
\end{figure}

Figure~\ref{fig:fig1} presents the temperature-dependent (a) and concentration-dependent (b) variations in the density of magnetic fluids. The temperature dependencies (Figure~\ref{fig:fig1}a) exhibit a linear decrease in density with rising temperature, which is attributed to the thermal expansion of the carrier medium. The concentration dependencies (Figure~\ref{fig:fig1}b) show that density increases with the volume fraction of magnetite nanoparticles, reflecting the higher density of magnetite compared to the carrier fluid. Solid lines in Figure 1b represent calculated values based on a linear density mixing model \(\rho = \rho_{2}\varphi + \rho_{1}(1-\varphi)\). The excellent agreement between experimental and calculated data confirms the accuracy of the reported magnetite concentrations and their consistency with expected values.

\begin{figure}[h!]
\centering
\includegraphics[width=0.75\linewidth]{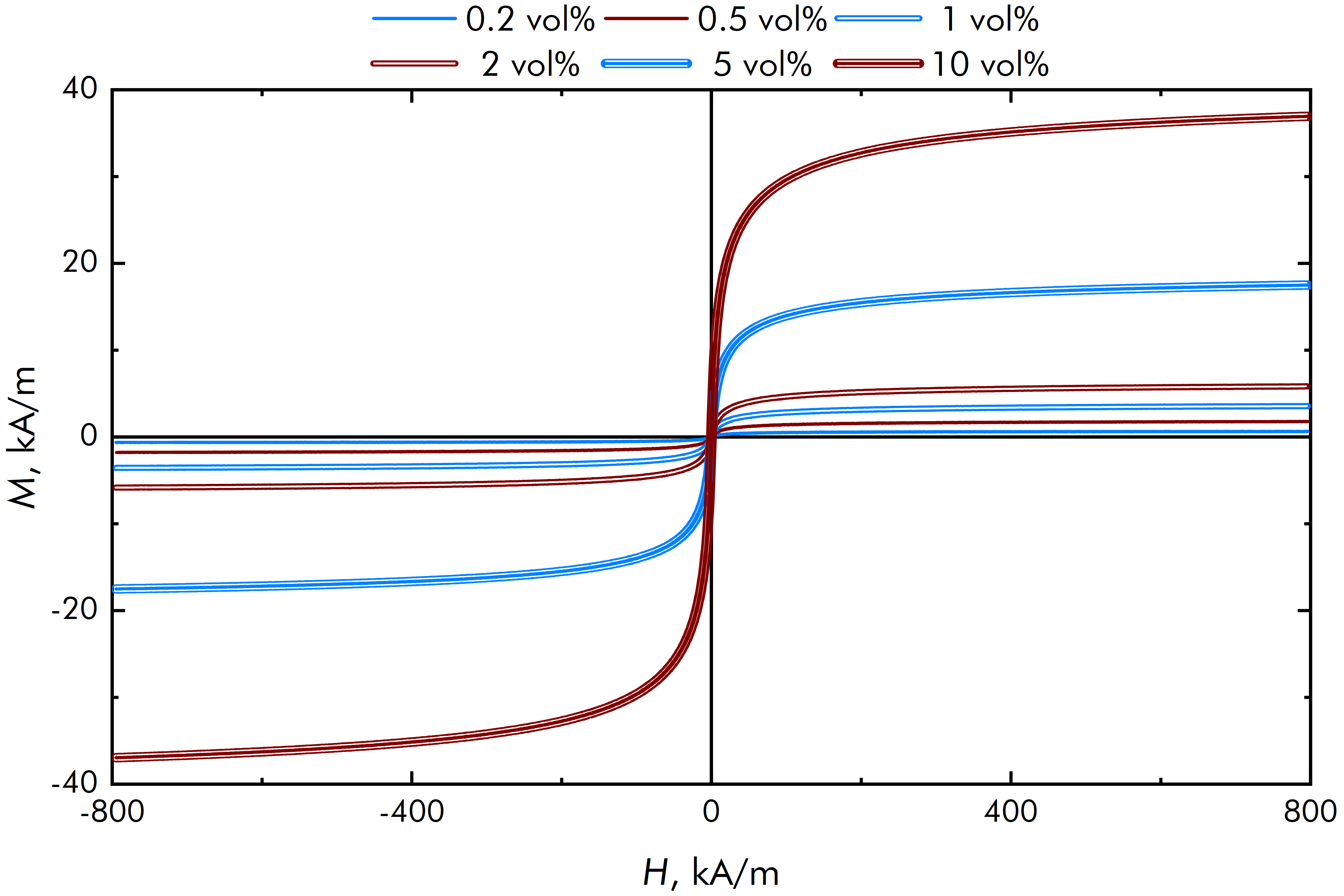}
\caption{Magnetization curves of magnetic fluid samples with varying nanoparticle concentrations.}
\label{fig:fig2}
\end{figure}

Figure~\ref{fig:fig2} shows the magnetization curves of the magnetic fluid samples measured at room temperature. The curves exhibit typical superparamagnetic behavior with negligible hysteresis and an almost zero coercive force, indicating that the magnetic core size of the nanoparticles does not exceed 25~nm \cite{jeun2012physical}. Magnetization curve analysis was used to determine the initial magnetic susceptibility (\(\chi_L\)). For dilute samples, \(\chi_L\) increases linearly with the nanoparticle volume fraction (\(\varphi\)), indicating minimal interparticle interactions. However, in concentrated systems, dipole-dipole interactions become significant and are described by a second-order modified mean-field theory \cite{solovyova2020initial, ivanov2020dynamic}. This model provides the best fit to the experimental data and allows estimation of the interparticle interaction parameter (\(\lambda\)), which was determined to be \(\lambda = 15.4\) (SI units).

\begin{figure}[h!]
\centering
\includegraphics[width=0.75\linewidth]{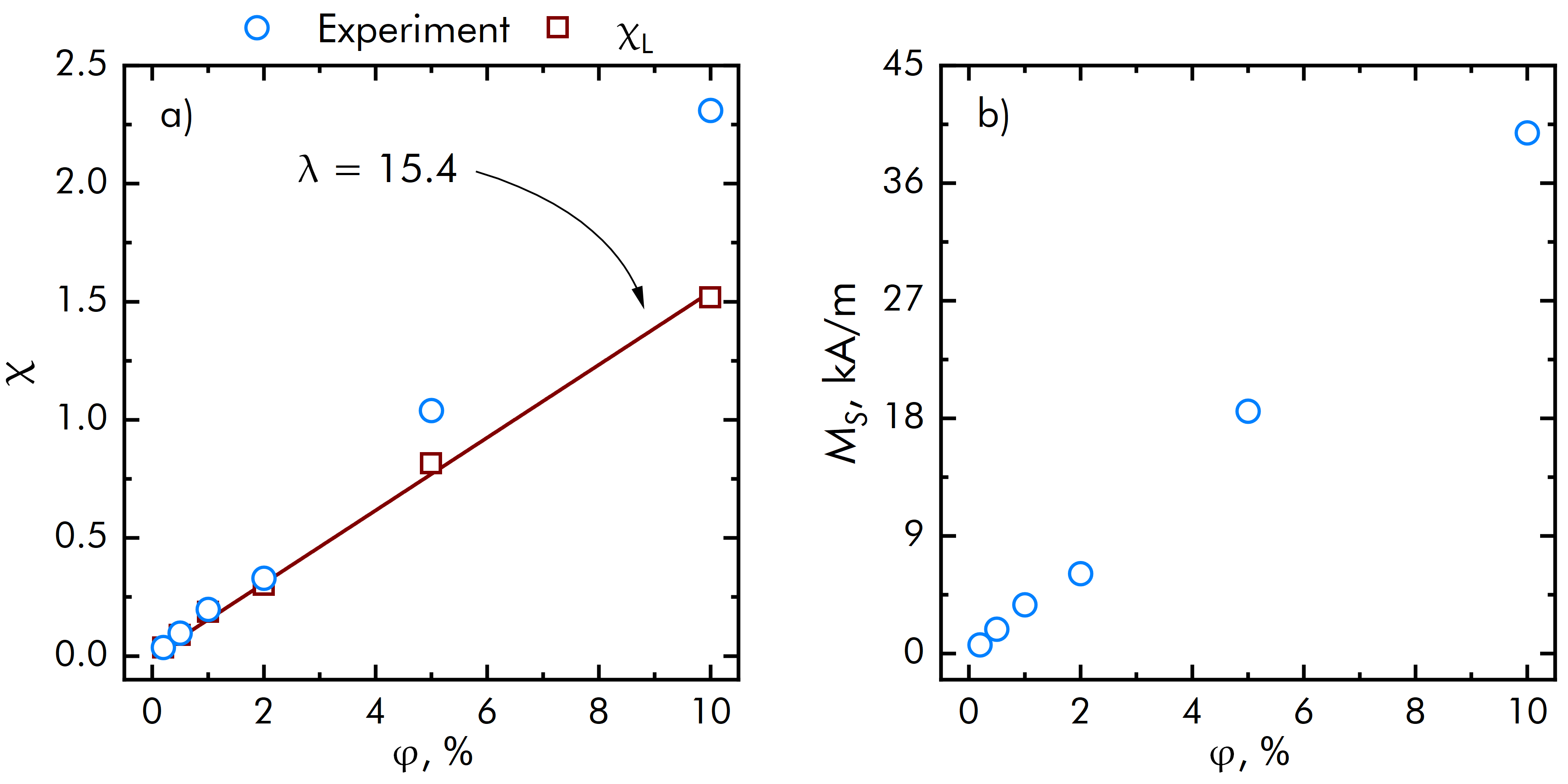}
\caption{Concentration dependencies of initial magnetic susceptibility (a) and saturation magnetization (b) for magnetic fluid samples.}
\label{fig:fig3}
\end{figure}

Figure~\ref{fig:fig3} illustrates the concentration-dependent initial magnetic susceptibility (\(\chi_L\)) (a) and saturation magnetization (\(M_S\)) (b). The concentration dependence of \(\chi_L\) (Figure~\ref{fig:fig3}a) confirms a linear increase up to 2~vol\%, indicating weak interparticle interactions. Beyond this concentration, \(\chi_L\) exhibits an additional increase consistent with the modified mean-field theory. Figure~\ref{fig:fig3}b reveals that the saturation magnetization of the nanoparticles is approximately 20\% lower than that of bulk magnetite (\(480~\text{kA/m}\)). This reduction is attributed to surface effects, such as spin frustration and the presence of defects on the nanoparticle surfaces \cite{lak2021embracing, singh2016defects}.

\begin{figure}[h!]
\centering
\includegraphics[width=0.75\linewidth]{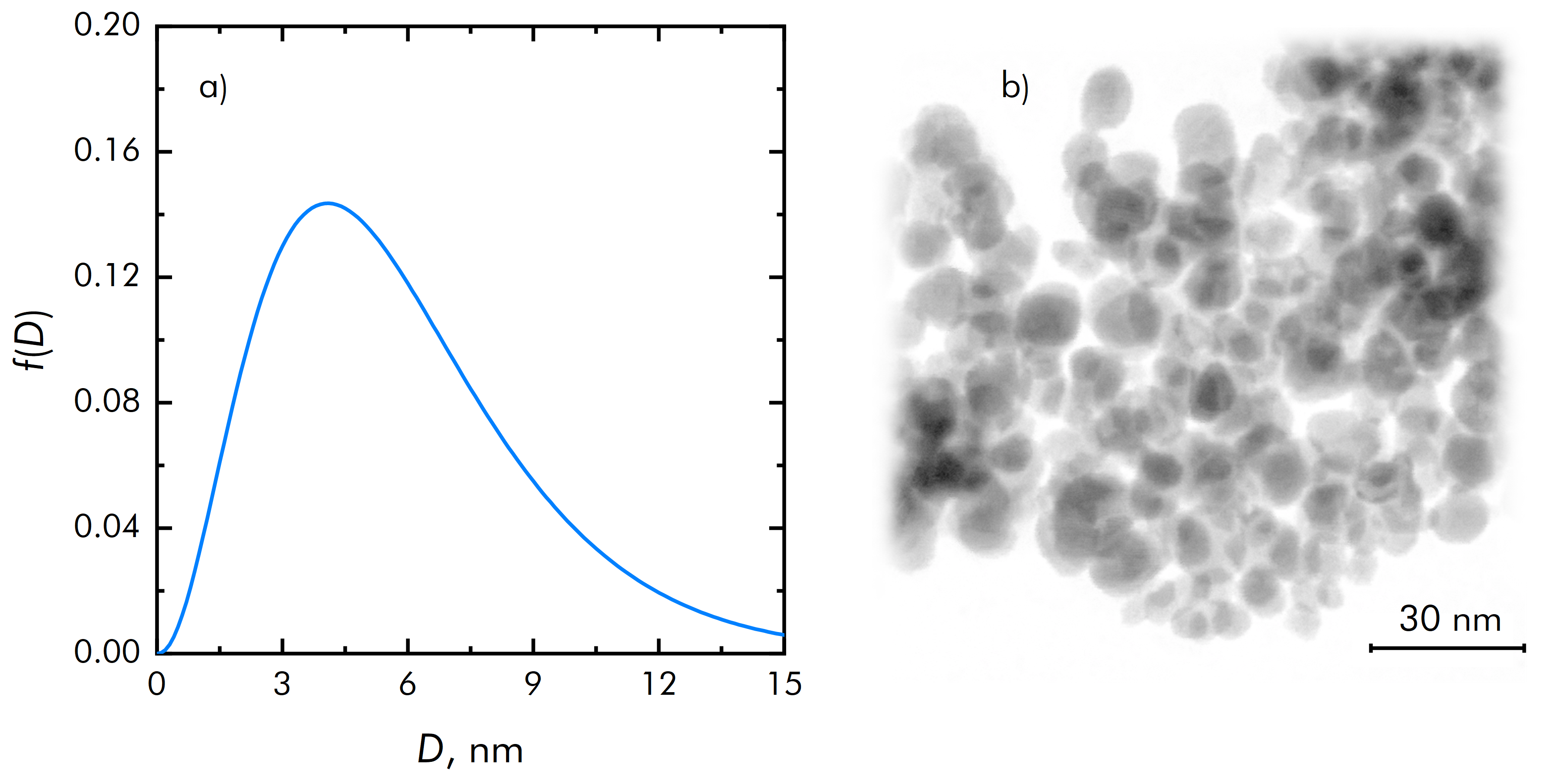}
\caption{Results of magnetic granulometric analysis (a) and TEM images of magnetite nanoparticles (b).}
\label{fig:fig4}
\end{figure}

Figure~\ref{fig:fig4} displays the particle size distribution analysis. Magnetic granulometric analysis (Figure~\ref{fig:fig4}a) determined the probability density function of magnetic core sizes, with an average size of 5~nm. Comparisons with transmission electron microscopy (TEM) images (Figure~\ref{fig:fig4}b) indicated larger nanoparticle sizes (approximately 16~nm). This discrepancy arises because TEM measures the total particle diameter, including any surfactant or aggregated layers, whereas magnetic granulometry reflects only the magnetic core size. Additionally, sample preparation steps for TEM, such as washing and sedimentation, may have contributed to the observed particle size increase due to aggregate formation.

\subsection{Dielectric Spectroscopy of Non-Magnetized Magnetic Fluids}
Dielectric spectroscopy is a powerful method for studying the electrical properties of materials, allowing the determination of parameters such as dielectric permittivity and conductivity, which play a critical role in understanding a system's behavior under an electric field.

\begin{figure}[h!]
\centering
\includegraphics[width=0.75\linewidth]{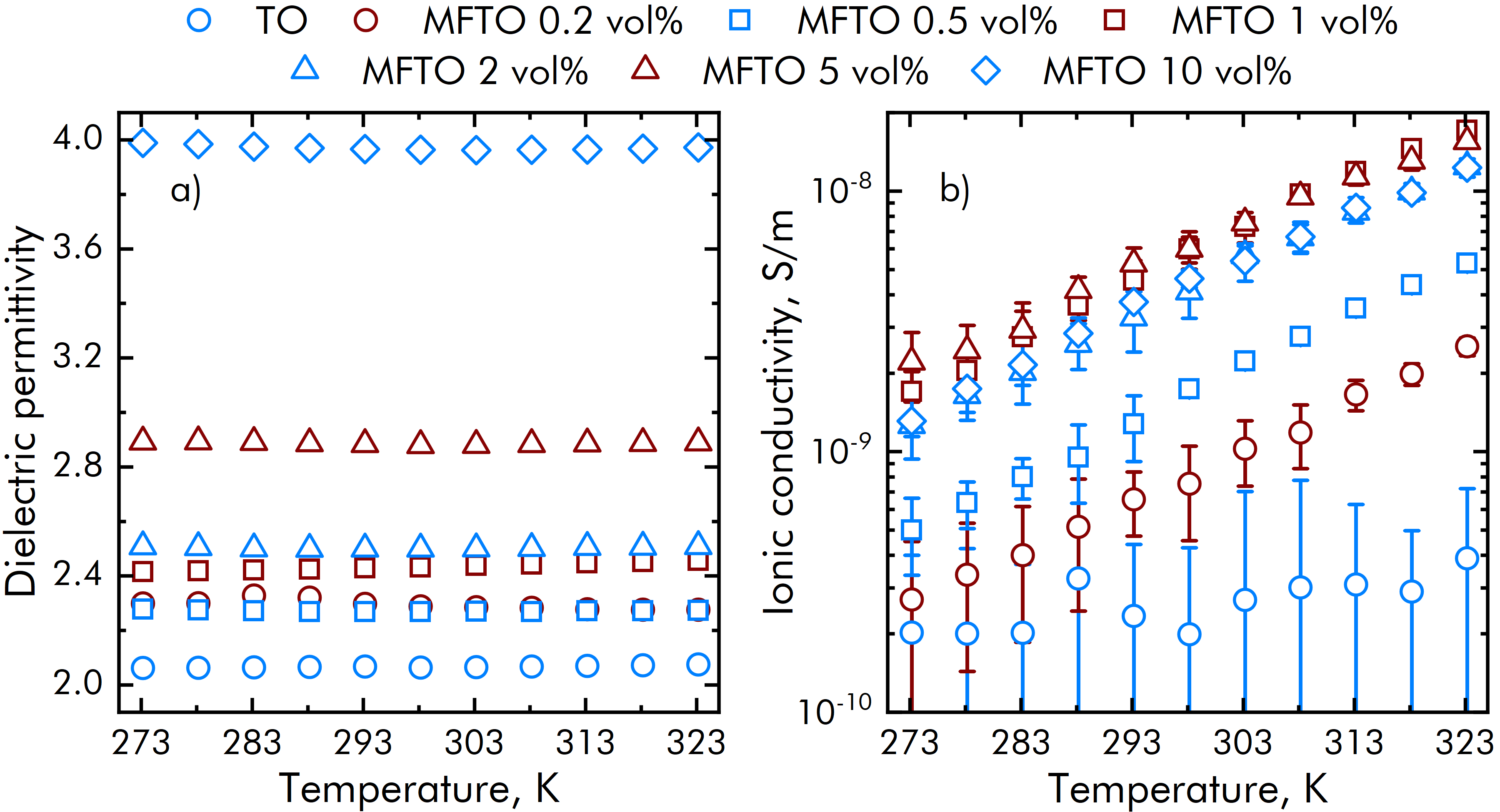}
\caption{Temperature dependencies of dielectric permittivity (a) and ionic conductivity (b) for magnetic fluid samples.}
\label{fig:fig5}
\end{figure}

Figure~\ref{fig:fig5} shows the temperature dependencies of dielectric permittivity (a) and ionic conductivity (b) for magnetic fluids with varying volume fractions of magnetite. Across all samples, the temperature dependence of dielectric permittivity is negligible. The carrier medium (transformer oil, TO) is characterized by minimal dielectric permittivity values (\(\sim 2.0\text{--}2.05\)), which increase significantly upon the addition of magnetite nanoparticles. At 10~vol\% magnetite, the permittivity reaches a maximum value (\(\sim 4.0\)), driven by nanoparticle-induced polarization and additional mechanisms such as interfacial polarization (Maxwell-Wagner mechanism \cite{adohi2008application, mansoorifar2017accuracy}) and the contribution of the electric double layer (EDL) surrounding the nanoparticles \cite{dong2017insight, joseph2023complex, zhao2006dielectric}.

\begin{figure}[h!]
\centering
\includegraphics[width=0.75\linewidth]{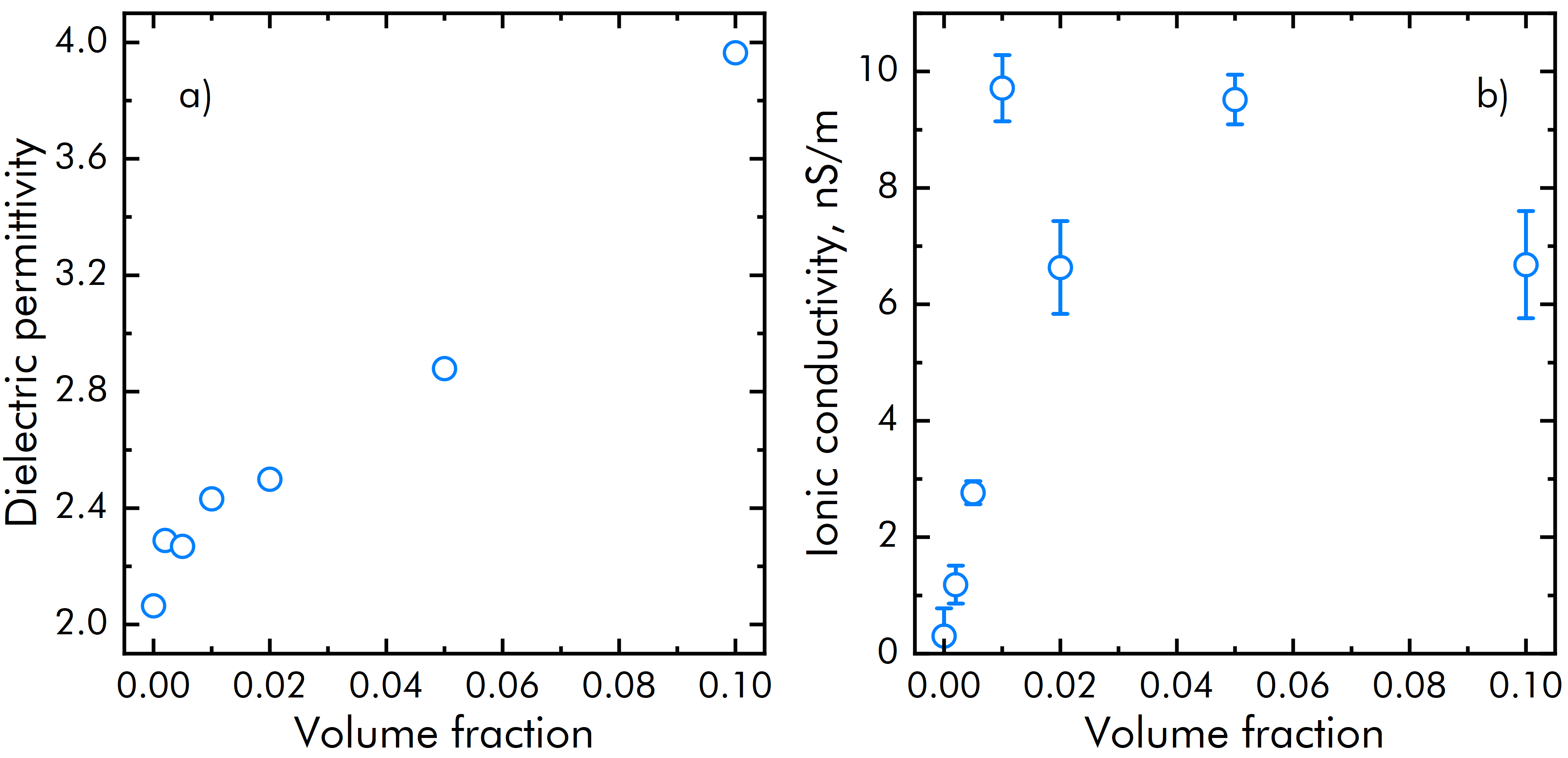}
\caption{Concentration dependencies of dielectric permittivity (a) and ionic conductivity (b) for magnetic fluid samples.}
\label{fig:fig6}
\end{figure}

The ionic conductivity of all investigated systems increases with temperature due to the enhanced kinetic energy of ions, which facilitates diffusion. The pure carrier medium exhibits minimal ionic conductivity (\(\sim 0.2\text{--}0.3\,\text{nS/m}\)), which increases gradually with temperature. The introduction of magnetite nanoparticles causes a sharp rise in conductivity; at 1~vol\%, conductivity increases 10–60 times, depending on temperature. This behavior is attributed to the formation of additional charge transfer pathways via ions adsorbed on nanoparticle surfaces. However, further increases in magnetite concentration lead to saturation or even a decline in conductivity. For samples containing 1–10~vol\% magnetite, conductivity stabilizes at \(7\text{--}10\,\text{nS/m}\) at room temperature. This decrease may result from ion trapping by nanoparticles, whose surfactant-coated surfaces participate in ion exchange or create localized ion traps \cite{zhakin2023study}. Concentration-dependent dielectric permittivity and ionic conductivity at room temperature are shown in Figure~\ref{fig:fig6} for clarity.

\begin{figure}[h!]
\centering
\includegraphics[width=0.75\linewidth]{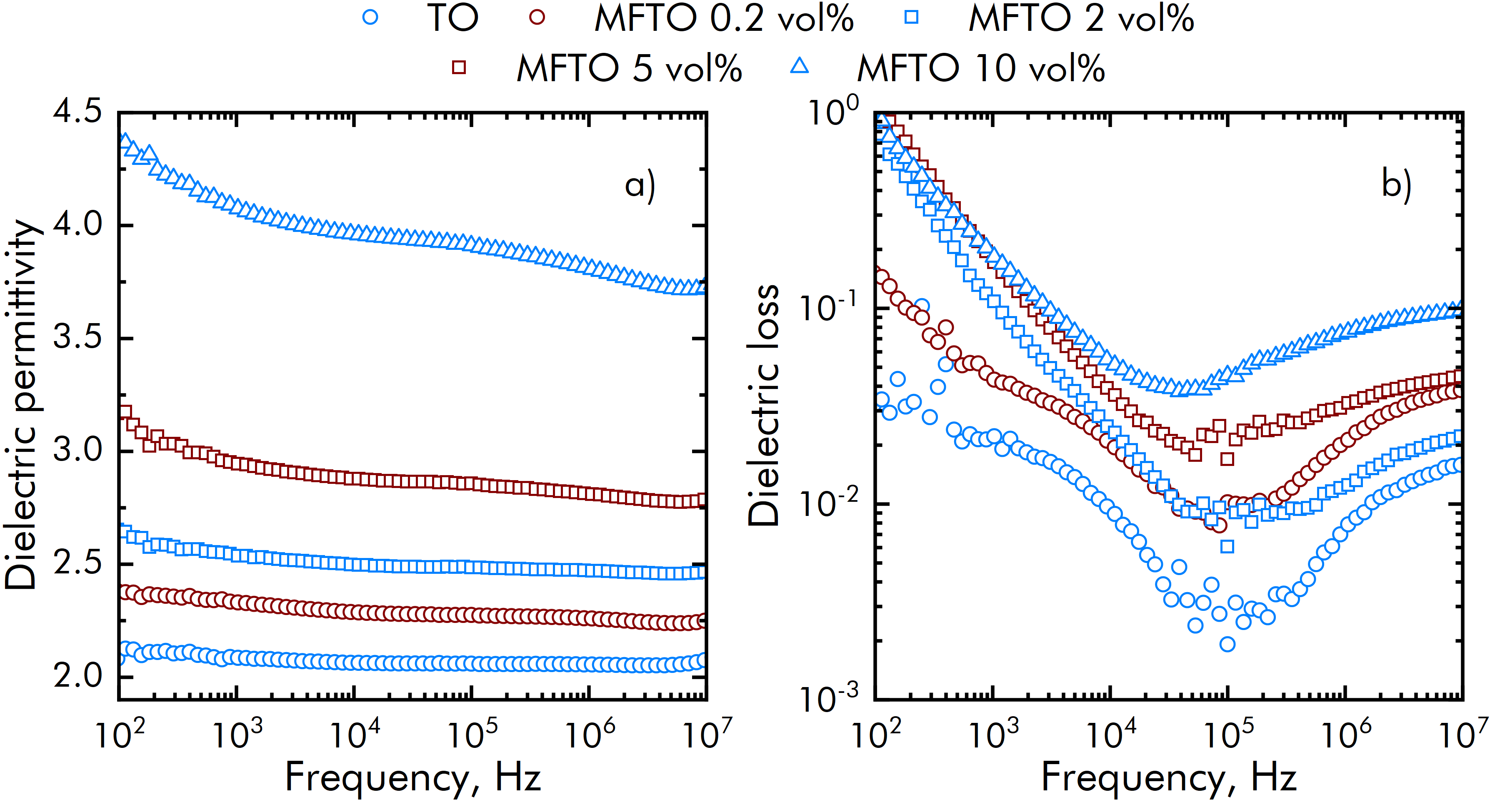}
\caption{Spectra of the real (a) and imaginary (b) components of dielectric permittivity for magnetic fluid samples.}
\label{fig:fig7}
\end{figure}

Figure~\ref{fig:fig7} depicts the spectra of the real (a) and imaginary (b) components of dielectric permittivity for non-magnetized magnetic fluids at \(25\,^{\circ}\text{C}\). Two characteristic dispersion regions are observed: a low-frequency region (below \(10^{3}\,\text{Hz}\)) and a high-frequency region (\(10^{5}\text{--}10^{7}\,\text{Hz}\)).

In the low-frequency region, the primary mechanism is the polarization of the electric double layer. This effect manifests as a characteristic increase in dielectric permittivity and losses with increasing magnetite concentration, caused by charge accumulation at phase boundaries facilitated by the EDL around nanoparticles. Consequently, EDL polarization intensifies with the number of nanoparticles.

In the intermediate and high-frequency regions (\(10^{5}\text{--}10^{7}\,\text{Hz}\)), pronounced dielectric dispersion is accompanied by significant increases in dielectric losses. This behavior is attributed to relaxation processes associated with interfacial polarization of nanoparticle aggregates. At high nanoparticle concentrations, the dispersion spectrum becomes broader, indicating a wide range of relaxation times. This phenomenon is well described by the Havriliak-Negami model \cite{gorska2018havriliak}, which accounts for the distribution of characteristic relaxation times in complex multicomponent systems.

The broadening of the spectrum arises from aggregation processes, where complex structures with varied relaxation times are formed. Interactions among particles within aggregates lead to significant variability in relaxation times and a widening of the dispersion spectrum. The observed increase in dielectric losses at high magnetite concentrations confirms that the primary contribution to dispersion arises from charge relaxation at aggregate-matrix interfaces.

For individual magnetite nanoparticles in transformer oil, Maxwell-Wagner relaxation occurs in the GHz range \cite{fannin2021magnetically, marin2024polarizing}, excluding its contribution to the observed \(10^{5}\text{--}10^{7}\,\text{Hz}\) dispersion. However, in aggregates, the characteristic relaxation time increases due to collective charge accumulation effects and weakened interaction with the surrounding medium. This shifts the relaxation frequency to the range corresponding to the observed dispersion. Thus, at high magnetite concentrations, aggregates dominate the dielectric properties of the magnetic fluid.

\begin{figure}[h!]
\centering
\includegraphics[width=0.75\linewidth]{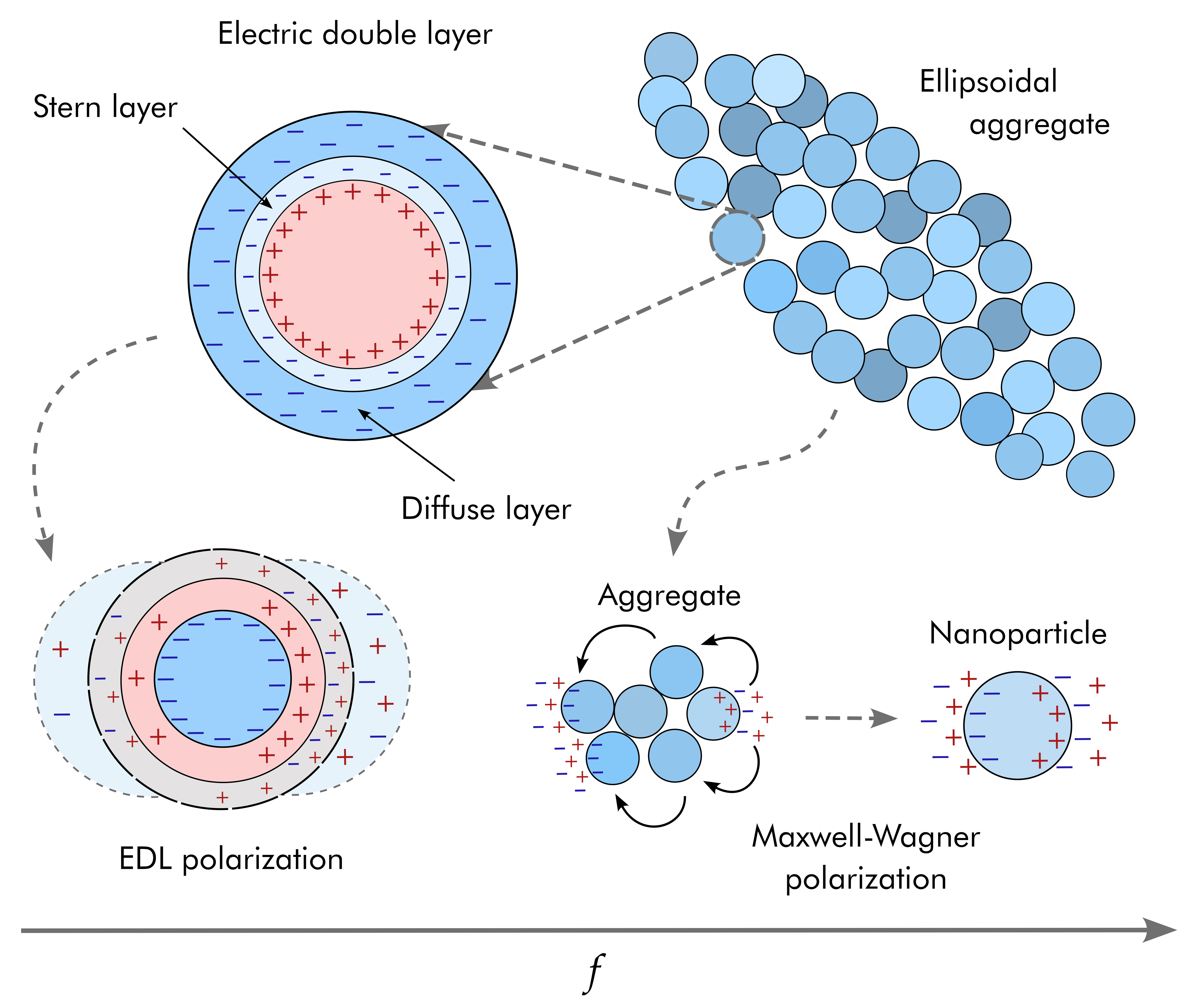}
\caption{Schematic representation of polarization mechanisms in magnetic fluids with ellipsoidal aggregates.}
\label{fig:fig8}
\end{figure}

Similar polarization mechanisms are observed in magnetized magnetic fluids. These include EDL polarization, interfacial polarization of anisometric aggregates, and single nanoparticle polarization. These processes underlie the dispersive behavior and unique properties of magnetic fluids in variable magnetic fields. A schematic representation of the polarization mechanisms for magnetic fluids containing ellipsoidal aggregates is shown in Figure~\ref{fig:fig8}.

\subsection{Dielectric Spectroscopy of Magnetized Magnetic Fluids}
\begin{figure}[h!]
\centering
\includegraphics[width=0.75\linewidth]{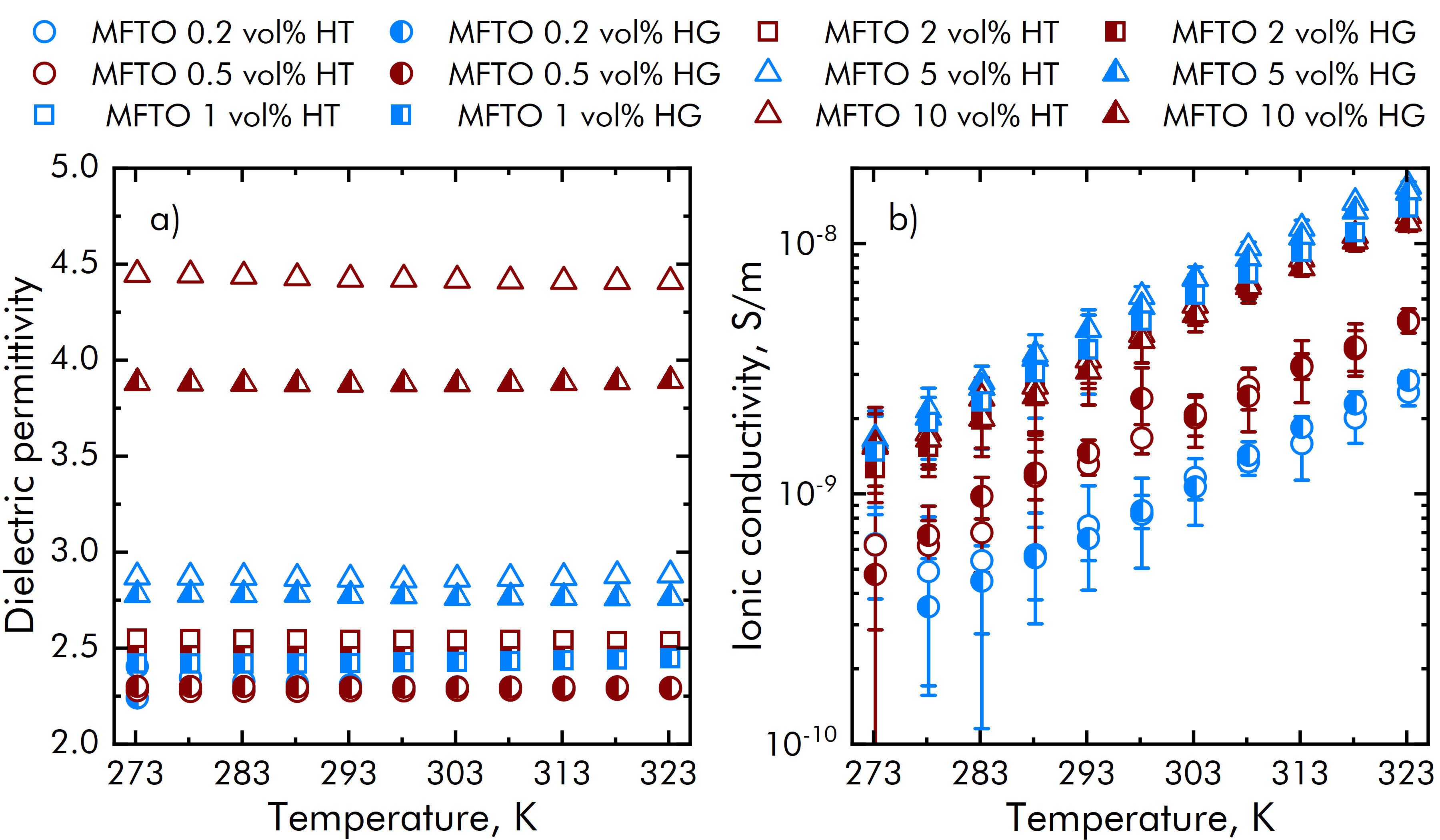}
\caption{Temperature dependencies of dielectric permittivity (a) and ionic conductivity (b) for magnetic fluid samples under parallel (HT) and perpendicular (HG) magnetic field orientations relative to the electric field.}
\label{fig:fig9}
\end{figure}

The temperature dependencies of dielectric permittivity and ionic conductivity for magnetized magnetic fluids under parallel (HT) and perpendicular (HG) magnetic field orientations are shown in Figures~\ref{fig:fig9}a and ~\ref{fig:fig9}b, respectively. Similar to non-magnetized fluids, the dielectric permittivity exhibits negligible temperature dependence. However, its values vary significantly with the orientation of the magnetic field. In the parallel orientation (HT), the dielectric permittivity is substantially higher than in the perpendicular orientation (HG). This difference arises from the formation of elongated nanoparticle aggregates aligned along the magnetic field, enhancing the contribution of interfacial polarization in the parallel configuration \cite{sokolov2010wave, kurilov2023concentration}.

Ionic conductivity, on the other hand, increases with temperature, showing relatively small differences between the parallel and perpendicular orientations. This behavior can be attributed to the limited impact of aggregate orientation on ion migration, unlike its more pronounced influence on the system's polarization properties.

\begin{figure}[h!]
\centering
\includegraphics[width=0.75\linewidth]{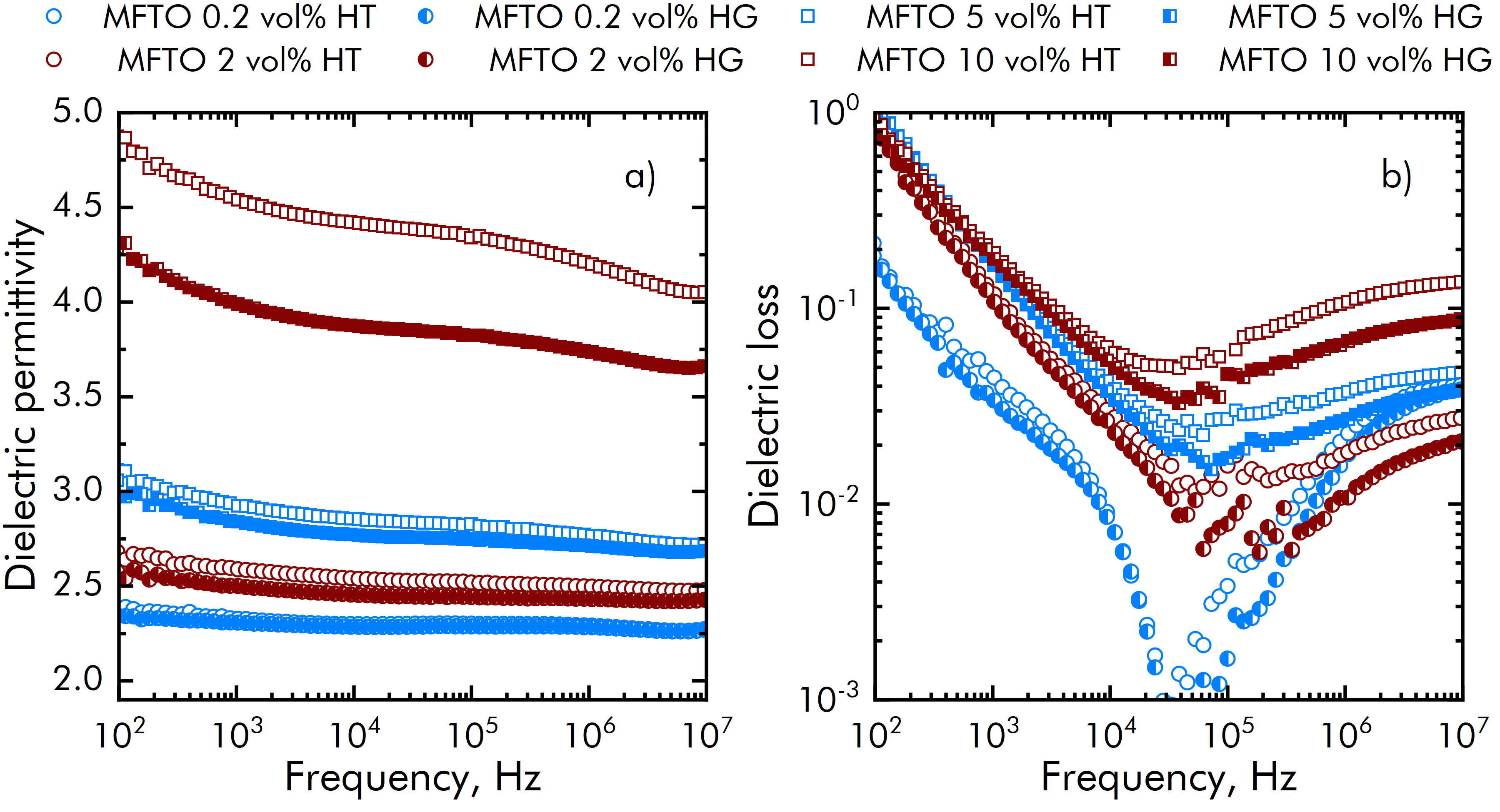}
\caption{Spectra of the real (a) and imaginary (b) components of dielectric permittivity for magnetic fluid samples under parallel (HT) and perpendicular (HG) magnetic field orientations relative to the electric field.}
\label{fig:fig10}
\end{figure}

Dielectric spectra of magnetized magnetic fluids with varying nanoparticle concentrations and magnetic field orientations, measured at \(25\,^{\circ}\text{C}\), are shown in Figure~\ref{fig:fig10}. The primary differences between non-magnetized and magnetized fluids lie in the appearance of dielectric anisotropy and variations in the strength of dielectric relaxation.

Dielectric anisotropy, observed as the difference in permittivity between parallel and perpendicular orientations, exhibits a clear frequency dependence: it decreases with increasing frequency. This behavior is attributed to the diminishing contribution of interfacial polarization from elongated aggregates at higher frequencies.

Dielectric losses, shown in Figure~\ref{fig:fig10}b, also depend on the magnetic field orientation. In the frequency range of \(10^{5}\text{--}10^{7}\,\text{Hz}\), losses in the parallel orientation significantly exceed those in the perpendicular orientation, reflecting stronger relaxation processes under parallel alignment. This is due to the higher effective polarization of aggregates when the magnetic field is parallel to the electric field.

\begin{figure}[h!]
\centering
\includegraphics[width=0.75\linewidth]{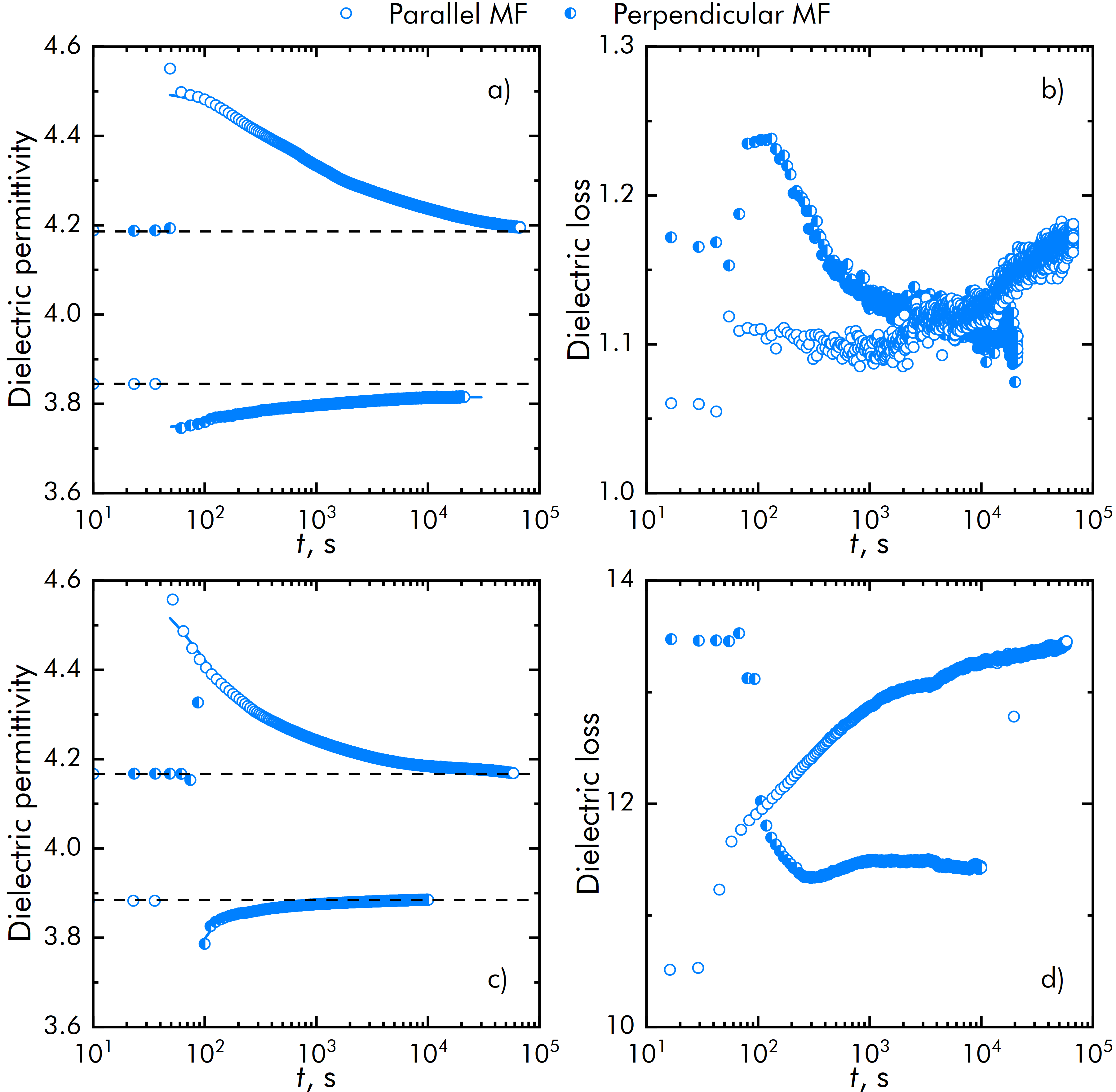}
\caption{Time-dependent dielectric permittivity (a, c) and dielectric losses (b, d) for magnetic fluids under changes in magnetic field orientation at different temperatures: \(0\,^{\circ}\text{C}\) (a, b) and \(50\,^{\circ}\text{C}\) (c, d).}
\label{fig:fig11}
\end{figure}

Temporal changes in dielectric permittivity (10~kHz) and dielectric losses (20~Hz) of a magnetic fluid with a 10~vol\% magnetite concentration under abrupt switching of the magnetic field orientation (from parallel to perpendicular and vice versa) are presented in Figure~\ref{fig:fig11}. Data are shown for two temperatures: \(0\,^{\circ}\text{C}\) (Figures~\ref{fig:fig11}a and~\ref{fig:fig11}b) and \(50\,^{\circ}\text{C}\) (Figures~\ref{fig:fig11}c and~\ref{fig:fig11}d). These results illustrate the dynamics of aggregate reorientation and structural evolution under the influence of the magnetic field. The formation of the internal structure of magnetic fluids is governed by diffusion processes \cite{sokolov2010wave}, which can require significant time, often up to tens of hours.

The time-dependent permittivity curves are well-approximated by a two-phase exponential decay function with a temporal offset, as indicated by solid lines on the graphs. At \(0\,^{\circ}\text{C}\), the characteristic relaxation times are \(t_{1} = 510\,\text{s}\) and \(t_{2} = 6140\,\text{s}\), reflecting the relatively slow restructuring of aggregates due to the high viscosity of the fluid. At \(50\,^{\circ}\text{C}\), the formation of a stable structure is significantly accelerated, with relaxation times of \(t_{1} = 98\,\text{s}\) and \(t_{2} = 1620\,\text{s}\). This acceleration is attributed to the reduced viscosity of the carrier medium at elevated temperatures, facilitating nanoparticle aggregate reconfiguration in the magnetic field.

\subsection{Discussion of Results}
\begin{figure}[h!]
\centering
\includegraphics[width=0.75\linewidth]{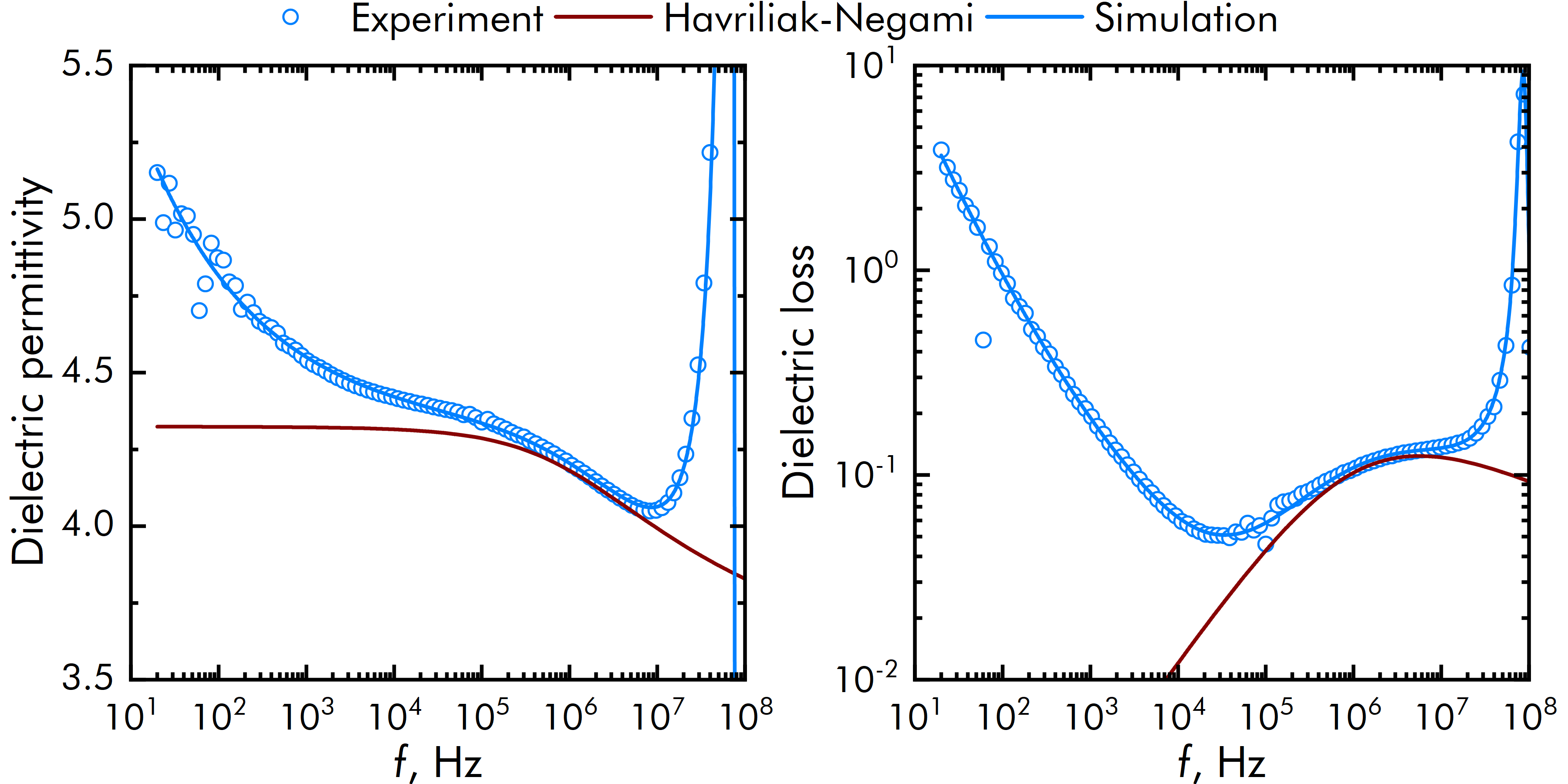}
\caption{Dielectric spectrum of a magnetic fluid at \(25\,^{\circ}\text{C}\): experimental data and modeling with relaxation and additional contributions.}
\label{fig:fig12}
\end{figure}

Figure~\ref{fig:fig12} presents the dielectric spectrum of a magnetic fluid with 10 vol\% magnetite concentration at \(25\,^{\circ}\text{C}\), approximated using the generalized Havriliak-Negami model. The solid red line represents the approximation results, accounting for relaxation time distributions and asymmetry in the dispersion curve \cite{gorska2018havriliak}. According to this model, the dielectric spectrum is described by the equation
\begin{equation*}
	\varepsilon^{*} = \varepsilon' - i\varepsilon'' = \varepsilon_{\infty} + \dfrac{\delta\varepsilon}{\left[1 + \left(i\omega\tau\right)^\alpha\right]^{\beta}},
\end{equation*}
where \(\varepsilon_{\infty}\) is the high-frequency limit of dielectric permittivity, \(\delta\varepsilon = \varepsilon_S - \varepsilon_{\infty}\) is the relaxation strength, \(\varepsilon_S\) is the static (low-frequency) permittivity, \(\tau\) is the relaxation time, \(\alpha\) specifies the relaxation time distribution's width, while \(\beta\) accounts for the dispersion curve's asymmetry.

To accurately describe the experimental data, additional mechanisms that significantly influence the dielectric spectrum were considered. At low frequencies (\(<10^2\,\text{Hz}\)), electrode polarization contributes substantially to the real part of the permittivity. This effect, proportional to \(A/f^k\), where \(A\) and \(k\) are empirical parameters, reflects charge accumulation at the electrode surfaces or ion migration processes. This empirical correction accounts for electrode polarization phenomenologically.

Additionally, ionic conductivity significantly affects dielectric loss dispersion at low frequencies. This behavior is described using the model
\begin{equation*}
	\varepsilon'' = \dfrac{\sigma_{DC}}{\omega\varepsilon_{0}} \left(1 + \left(\dfrac{\omega}{\omega_{p}}\right)^n\right),
\end{equation*}
where \(\sigma_{DC}\) is the static conductivity, \(\omega_p\) is the ionic hopping frequency, \(\varepsilon_0\) is the vacuum permittivity, and \(n\) characterizes the frequency dependence. In the absence of electrode polarization, (\(n = 1\), corresponding to ohmic conductivity.

At high frequencies (\(>10^6\,\text{Hz}\)), parasitic contributions from resonances in the measurement cell emerge, related to the cell’s inductance (\(L\)) and resistance (\(R\)). These effects were accounted for following methods outlined in previous studies \cite{perkowski2012dielectric, perkowski2021parasitic}.

Combining these factors yielded excellent agreement between experimental data and the theoretical model, shown by the solid blue line in Figure~\ref{fig:fig12}.

To further analyze the dielectric properties, Boyle’s model for polarizable systems with ellipsoidal particles was employed \cite{boyle1985electrical}. The effective permittivity \(\varepsilon^*\) of a suspension of oriented ellipsoids is given by \cite{asami2002characterization}
\begin{equation*}
	1 - \varphi = \left(\dfrac{\varepsilon^* - \varepsilon^*_2}{\varepsilon^*_1 - \varepsilon^*_2}\right)\left(\dfrac{\varepsilon^*_1}{\varepsilon^*}\right)^{L_k},
\end{equation*}
where \(\varepsilon^*_1\) and \(\varepsilon^*_2\) are the complex permittivities of the medium and aggregates, respectively, \(\varphi\) is the aggregate volume fraction, and \(L_k\) is the depolarization factor dependent on aggregate shape \cite{mejdoubi2006finite}.

In numerical modeling, the aggregate volume fraction was assumed equal to the magnetite concentration, while the effective permittivities of the aggregates and medium, along with the aspect ratio of ellipsoids, were treated as variables. The results validated the model's suitability, enabling detailed descriptions of relaxation processes in magnetic fluids.

\begin{figure}[h!]
\centering
\includegraphics[width=0.75\linewidth]{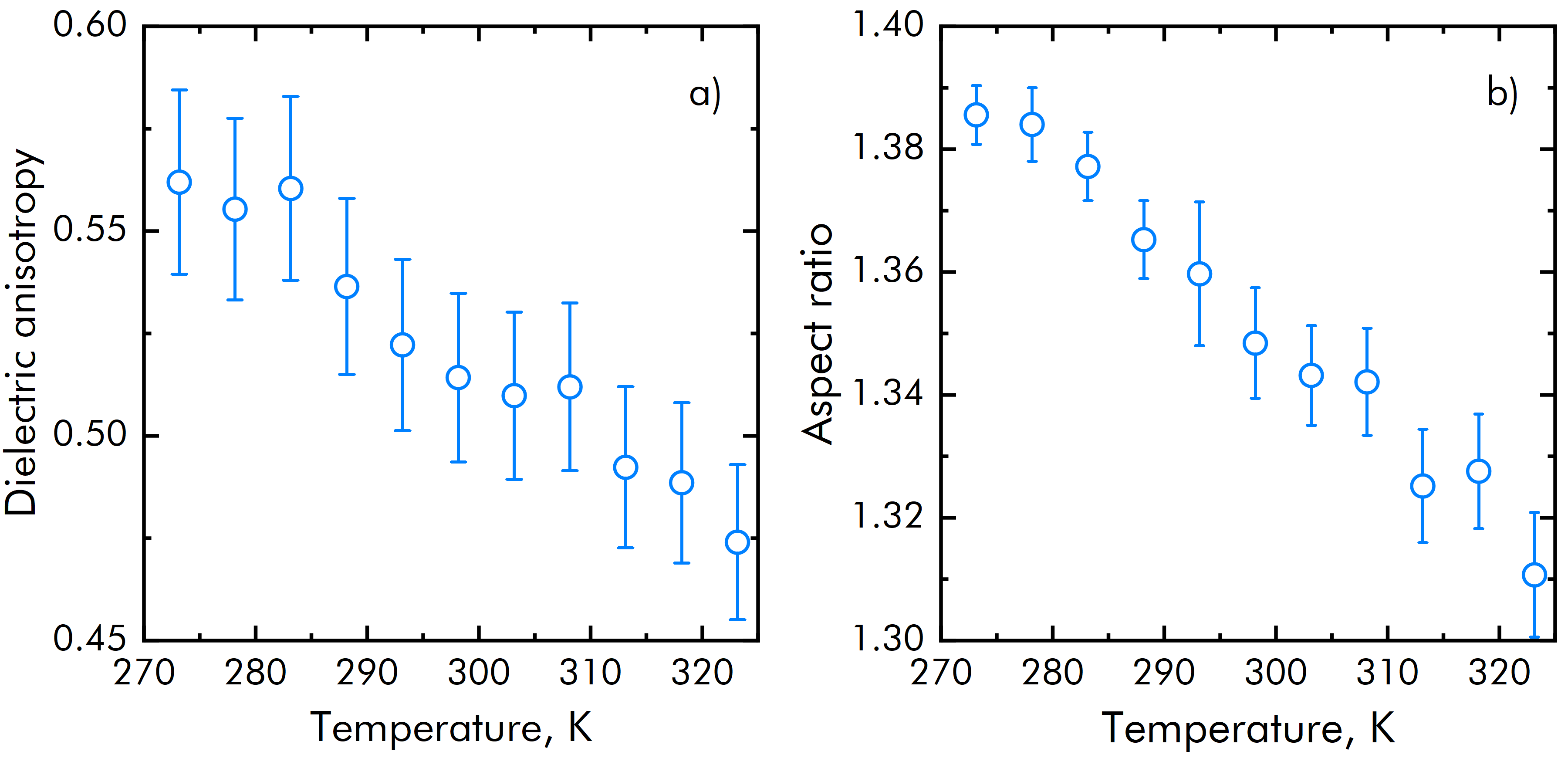}
\caption{Temperature dependencies of dielectric anisotropy (a) and aggregate aspect ratio (b) in magnetized magnetic fluids.}
\label{fig:fig13}
\end{figure}

Figure~\ref{fig:fig13} shows the temperature dependencies of dielectric anisotropy (a) and the aspect ratio of magnetically induced aggregates (b) obtained from the dielectric spectra of magnetic fluids with 10~vol\% magnetite. Dielectric anisotropy, representing the difference in the system’s response to an external electric field along different directions, indicates the internal structural order. The analysis reveals a decrease in anisotropy with increasing temperature, attributed to the reduction of the interparticle interaction parameter (\(\lambda\)), which leads to a loss of structural order in magnetically induced aggregates. This trend is supported by the decreasing aspect ratio of aggregates, indicating a transition from elongated structures to more compact forms.

\begin{figure}[h!]
\centering
\includegraphics[width=0.75\linewidth]{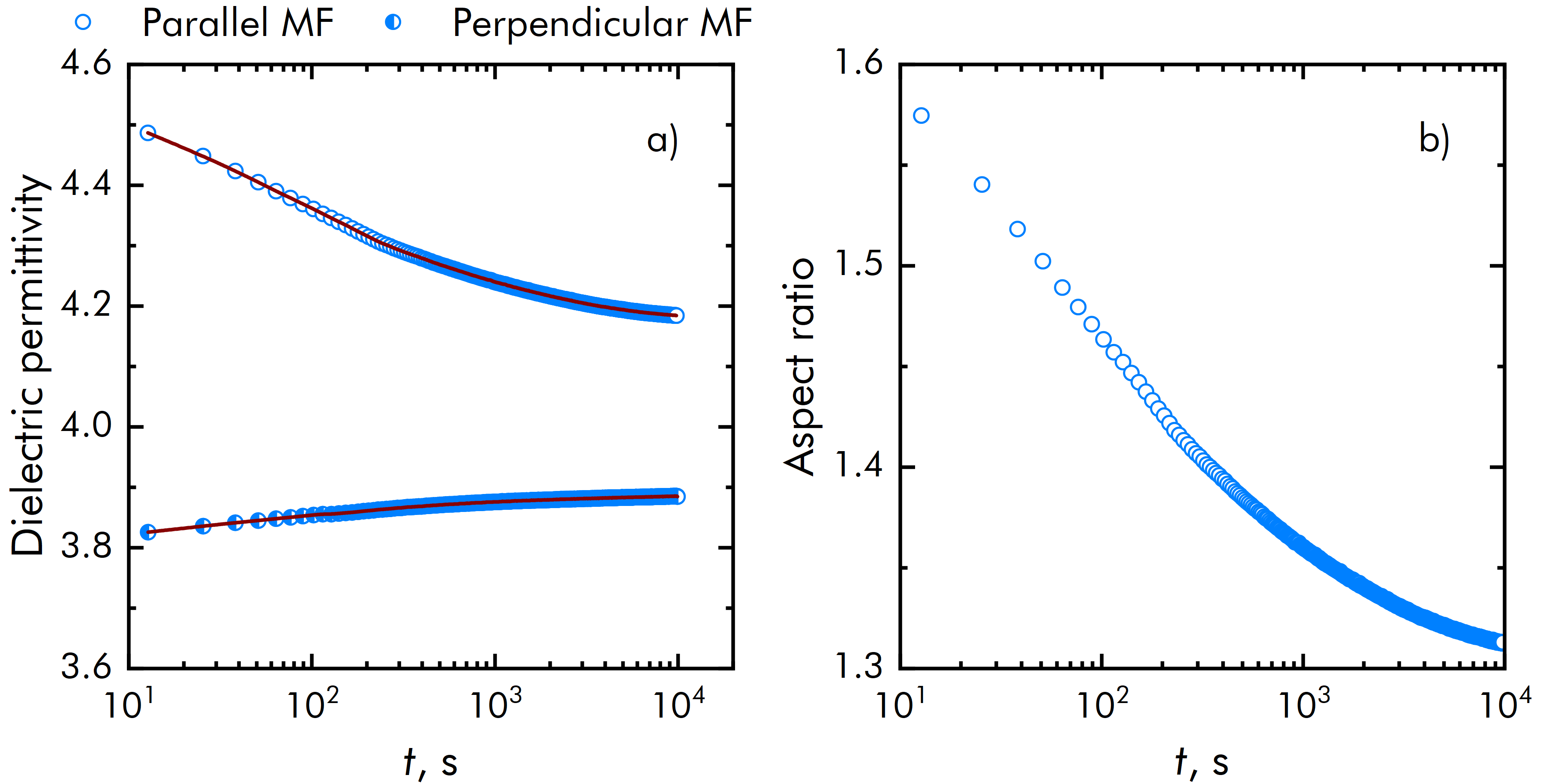}
\caption{Time-dependent dielectric permittivity (a) and aggregate aspect ratio (b) in magnetic fluids under magnetic field orientation changes at \(50\,^{\circ}\text{C}\).}
\label{fig:fig14}
\end{figure}

Figure~\ref{fig:fig14}a presents time-dependent dielectric permittivity (10~kHz) analyzed using the Boyle model. The results, represented by solid lines, demonstrate the synchronous approximation for parallel and perpendicular orientations. The data indicate significant structural changes in the magnetic fluid upon abrupt changes in magnetic field orientation.

At the initial stages, elongated chain-like aggregates form rapidly (Figure~\ref{fig:fig14}b), subsequently thickening and reaching equilibrium over several hours. This process aligns with experimental observations and theoretical models describing chain-like structure formation and phase rearrangements in moderate magnetic fields \cite{ivanov2020chain, sokolov2021absorption, patel2009hydrodynamics}.

It is important to note that aggregate aspect ratio data are qualitative due to limitations in dynamic measurements, including the restricted frequency range, potentially affecting the accuracy of time-dependent approximations. Furthermore, aggregate size heterogeneity, confirmed by optical experiments on magnetorheological fluids, adds complexity to data interpretation \cite{borin2020targeted, borin2020characterization, nowak2017magnetoviscous}.

These findings confirm that changes in external magnetic fields significantly alter the dielectric and structural properties of magnetic fluids. This underscores the critical role of aggregation processes in accurately predicting the performance of magnetic fluid devices, particularly under dynamic field conditions.

\section{Conclusion}
This study provides a comprehensive investigation of the structural and dielectric properties of transformer oil-based magnetic fluids over a wide concentration range. Particular attention was given to the dynamic aggregation processes of nanoparticles under magnetic fields and their influence on the system’s electrophysical characteristics.

Dielectric spectroscopy revealed that increasing the concentration of magnetite nanoparticles leads to nonlinear changes in the effective dielectric permittivity. At low concentrations, both dielectric permittivity and ionic conductivity exhibit an almost linear increase, consistent with the Maxwell-Wagner model. At higher concentrations, effects such as conductivity saturation and dispersion in dielectric spectra emerge, driven by the formation of elongated aggregates. Analysis of the dielectric spectra demonstrated that the observed relaxation processes are caused by interfacial polarization of aggregates and charge accumulation in the double electrical layer.

Dielectric spectroscopy of magnetized samples showed that the orientation of the magnetic field significantly influences the internal structure and properties of magnetic fluids. The external magnetic field induces the formation of elongated structures, leading to frequency-dependent dielectric anisotropy. At higher frequencies, the contribution of interfacial polarization from elongated aggregates diminishes, reducing the dielectric anisotropy.

The results were analyzed using Boyle’s polarization model, which provided deeper insights into microstructure formation, relaxation processes, and dielectric anisotropy in high-concentration magnetic fluids. It was established that increasing temperature reduces aggregate ordering due to weakened interparticle interactions. This effect is corroborated by the decreasing aspect ratio of aggregates, indicating a transition from elongated aggregates to more compact structures.

Dynamic dielectric spectroscopy illustrated the complex reorganization dynamics of magnetically induced aggregates and their growth evolution under the influence of magnetic fields. Data analysis, based on Boyle’s model, shows that abrupt changes in magnetic field orientation induce significant alterations in the internal structure of magnetic fluids. At initial stages, elongated chain-like aggregates form rapidly and then thicken, reaching equilibrium over tens of hours.

This study advances the understanding of relaxation mechanisms, interfacial polarization, and aggregation dynamics in magnetic fluids. The findings have important implications for the design of high-performance devices operating in dynamic magnetic and electric fields, including sensors, magnetorheological devices, and electronic components.

\section*{Acknowledgments}
This research was supported by the Russian Science Foundation grant No. 24-29-00178 (\url{https://rscf.ru/project/24-29-00178/}).

\bibliographystyle{plain}
\bibliography{DeoLib}

\end{document}